\documentclass[aps,10pt,twocolumn,showpacs,amsmath,amssymb]{revtex4-1}

\usepackage{graphicx}
\usepackage{color}
\usepackage{soul}

\newcommand{\C}[1]{{\cal C}_{#1}}

\begin{document}


\title{Are we overlooking Lepton Flavour Universal New Physics in $b\to s\ell\ell$ ?}

%

\author{Marcel Alguer\'o$^{1,2}$, Bernat Capdevila$^{1,2}$, S\'ebastien Descotes-Genon$^{3}$, Pere Masjuan$^{1,2}$ and Joaquim Matias$^{1,2}$
\vspace{0.3cm}}

\affiliation{
$^{1}$ Grup de F\'isica Te\`orica (Departament de F\'isica), Universitat Aut\`onoma de Barcelona, E-08193 Bellaterra (Barcelona), Spain\\
$^{2}$ Institut de F\'isica d'Altes Energies (IFAE), The Barcelona Institute of Science and Technology, Campus UAB, E-08193 Bellaterra (Barcelona), Spain\\
$^{3}$Laboratoire de Physique Th\'eorique, UMR 8627, 
CNRS, Univ. Paris-Sud, Universit\'e Paris-Saclay, 91405 Orsay Cedex, France\\
}

\begin{abstract}
The deviations with respect to the Standard Model (SM) that are currently observed in $b \to s \ell\ell$ transitions (the so-called flavour anomalies) can be interpreted in terms of different New Physics (NP) scenarios within a model-independent effective approach. We reconsider the determination of NP in global fits from a different perspective by removing one implicit hypothesis of current analyses, namely that NP is only Lepton-Flavour Universality Violating (LFUV). We examine the roles played by LFUV NP and Lepton-Flavour Universal (LFU) NP altogether, providing new directions to identify the possible theory beyond the SM responsible for the anomalies observed. New patterns of NP emerge due to the possibility of allowing at the same time large LFUV and LFU NP contributions to $\C{10\mu}$, which provides a different mechanism to obey the 
constraint from the $B_s \to\mu^+\mu^-$ branching ratio. In this landscape of NP, we discuss how to discriminate among these scenarios in the short term thanks to current and forthcoming observables.
While the update of $R_K$ will be a major milestone to confirm the NP origin of the flavour anomalies, additional observables, in particular the LFUV angular observable $Q_5$, turn out to be central to assessing the precise NP scenario responsible for the observed anomalies.
\end{abstract}


\maketitle

\section{Introduction} 

Besides the fundamental discovery of the SM Higgs the first run at LHC had two clear outcomes. On  one side, no signals of NP have been found in direct searches. On the other side, indirect searches have led to a large set of deviations with respect to the SM (or anomalies) in both $b\to c\ell\nu$ and in $b\to s\ell\ell$ decays~\cite{Blake:2015tda,Bifani:2018zmi,Amhis:2016xyh}. We can classify the latter (which we focus on) in two sets: $b\to s \mu\mu$ anomalies related to observables testing only muonic transitions, which we call Lepton Flavour Dependent (LFD), and Lepton-Flavour Universality Violating (LFUV) anomalies that  correspond to deviations in observables comparing muonic and electronic transitions.

The $b\to s\ell\ell$ anomalies have been analysed in the effective Hamiltonian approach, which separate short- and long-distance contributions in a model-independent way (see, for instance, Ref.~\cite{Descotes-Genon:2015uva}). The analysis now combines the experimental data from LHC experiments (LHCb~\cite{Aaij:2013qta,Aaij:2015oid,Aaij:2013aln,Aaij:2014ora,Aaij:2017vbb} but also ATLAS~\cite{ATLAS:2017dlm} and CMS~\cite{CMS:2017ivg}) as well as the data from $B$-factories (in particular Belle~\cite{Abdesselam:2016llu,Wehle:2016yoi}) together with theoretical input concerning long-distance hadronic contributions~\cite{Khodjamirian:2010vf,Descotes-Genon:2014uoa,Capdevila:2017ert,Bobeth:2017vxj}. They aim at extracting the value of the short-distance Wilson coefficients under given NP hypotheses and at comparing them with the SM expectations. Even though different global analyses in the literature use different approaches (statistical treatment, observables, hadronic inputs\ldots), they agree on the emerging global picture~\cite{Descotes-Genon:2013wba,Capdevila:2017bsm,Altmannshofer:2017yso,Geng:2017svp,Ciuchini:2017mik,DAmico:2017mtc,Arbey:2018ics}. For instance, in Ref.~\cite{Capdevila:2017bsm}, a global fit including both LFD and LFUV observables finds pulls (comparing the statistical significance of the SM against that of a NP hypothesis) between 5.0 to 5.8$\sigma$, depending on the particular NP hypothesis used. The LFUV-NP hypotheses involving either $\C{9\mu}^{\rm NP}$ or $\C{9\mu}^{\rm NP}=-\C{10\mu}^{\rm NP}$ are among those with the highest significances.

In this letter we consider the possibility that short-distance Wilson coefficients will receive contributions from NP
that are not only LFUV but also Lepton Flavour Universal or LFU. Indeed, whereas LFUV-NP contributions are mandatory to explain $R_K$ and $R_{K^*}$,  $b\to s\ell\ell$ processes are not restricted to such NP contributions alone. While several articles~\cite{Capdevila:2017bsm,Altmannshofer:2017yso,Geng:2017svp,Ciuchini:2017mik,DAmico:2017mtc,Arbey:2018ics} allowed the presence of NP in electrons in global fits to $b\to s \ell\ell$, in the present paper we go one step beyond and we impose different types of LFU structures between all leptons. We show that a universal LFU-NP contribution, together with a LFUV-NP contribution, gives rise to scenarios with a statistical significance at least as relevant as the ones identified in Ref.~\cite{Capdevila:2017bsm}, against a common belief that the presence of such terms is not justified from the statistical point of view and should be dropped.
This may help to motivate the construction of new models including not only LFUV but also LFU-NP contributions.
Thus, we reconsider the results of the fit allowing for the presence of two different types of NP that may lead to a new paradigm concerning the nature of the underlying theory beyond the SM. We discuss then the next steps, to identify the NP scenario that is realized in Nature among the ones already favoured, complementing~\cite{Capdevila:2017bsm,Capdevila:2017ert}. 
Following our findings, the UV completion of the SM may require significant contributions from two different sectors (LFU and LFUV) instead of a single one as often assumed. 

\section{Two types of NP contributions}

The $b\to s\ell\ell$ processes can be analysed within the effective Hamiltonian framework~\cite{Buchalla:1995vs,Buras:1998raa}. The observables for exclusive decays can be written as interference terms between helicity amplitudes which are given
as (short-distance) Wilson coefficients multiplying (long-distance) hadronic matrix elements~\cite{Matias:2012xw,Descotes-Genon:2013vna,Altmannshofer:2008dz}, with a separation between short and long distance given by the factorisation scale $\mu_b=O(m_b)$. One can use the fact that $m_b$ is significantly larger than the typical QCD scale in order to isolate perturbatively computable contributions to the hadronic matrix elements (using effective approaches like QCD factorisation). These perturbative contributions of hadronic origin can be lumped together with the purely short-distance contribution into effective Wilson coefficients (that will multiply non-perturbative hadronic form factors) with the following structure in the case of $B\to K(^*)\ell\ell$~\cite{Beneke:2001at}:
\begin{align}\nonumber
\C{9\ell}(q^2)&=\C{9\, \rm pert}^{\rm SM}(q^2)+ \C{9}^{c\bar{c}}(q^2)+\C{9\ell}^{\rm NP} \\ 
\C{10\ell}&=\C{10}^{\rm SM}+\C{10\ell}^{\rm NP} \label{c9}
\end{align}
where $\ell = e,\mu$. The short-distance SM values \cite{Huber:2005ig} at this scale {$\mu_b=4.8\text{GeV}$} are $\C{9 }^{\rm SM}=4.07$ and 
$\C{10 }^{\rm SM}=-4.31$.
We have $\C{9 \, \rm pert}^{\rm SM}=\C{9 }^{\rm SM}+ Y(q^2)$, where the function $Y(q^2)$ stems from one-loop matrix elements of four-quark operators ${\cal O}_{1-6}$, corresponding to the $c\bar{c}$ continuum. It can be evaluated within perturbation theory at LO, and  corrections at ${\cal O}(\alpha_s)$ to $\C{9\ell}$ to this function are known~\cite{Buras:1994dj,Beneke:2001at,Greub:2008cy}. In addition to this continuum, there is a long-distance contribution, which corresponds in particular to charmonium resonances $\C{9}^{c\bar{c}}$ and depends on the external hadron state. Several approaches are available to estimate this contribution~\cite{Blake:2017fyh,Ciuchini:2017mik,Bobeth:2017vxj}, all with similar outcomes~\cite{Capdevila:2017ert,Arbey:2018ics}.
We follow here~\cite{Capdevila:2017ert,Descotes-Genon:2015uva}, using the light-cone sum rule computation with one soft-gluon exchange~\cite{Khodjamirian:2010vf} to get an order of magnitude estimate of this contribution, without making any assumption about its sign and thus allowing for constructive or destructive interference with the other contributions to $\C{9\mu}$.

This effective approach is the basis for global fits to the data in order to constrain the NP contributions $\C{i\ell}^{\rm NP}$ under various NP assumptions~\cite{Capdevila:2017bsm,Altmannshofer:2017yso,Geng:2017svp,Ciuchini:2017mik,Arbey:2018ics}. It turns out that the combination of anomalies in some LFD ($b\to s\mu\mu$) angular observables and in LFUV ratios $R_K$ and $R_{K^*}$ selects hypotheses with a large NP contribution to the Wilson coefficient $\C{9\mu}$ (of order 25\% of the SM),   or  NP contributions to both $\C{9\mu}$ and $\C{10\mu}$.

Following this perspective we ought to be more precise on what goes under the ``New Physics'' landscape. In this letter we  consider that the short-distance Wilson coefficients $\C{i\mu}$  can contain two types of NP contribution
\begin{equation}
\C{i\ell}^{\rm NP}=\C{i\ell}^{\rm V} +\C{i}^{\rm U} 
\end{equation}
with $\ell=e,\mu$ (the extension to $\tau$ is trivial, assuming true universality among $e$, $\mu$ and $\tau$) where $\C{i\ell}^{\rm V}$ stands for Lepton Flavour Universality Violating NP and $\C{i}^{\rm U}$ for Lepton Flavour Universal-NP contributions. 
These short-distance contributions are all independent of the external hadronic states and their kinematics; they differ therefore from long-distance hadronic contributions which are LFU, but dependent on the nature and kinematics of the hadronic states. We will define the separation between the two types of contributions by imposing that LFUV contributions affect only muons
\begin{equation} \C{ie}^{\rm V}=0
\end{equation}
There is no loss of generality here, since this term can always be absorbed in such a way that $\C{i\mu}^{\rm V}$ can be interpreted as the difference of NP contributions to muons and electrons.

\section{Global Fits in presence of LFU NP} 

LFUV-NP contributions are mandatory to explain LFUV anomalies. The $b\to s ee$ measurements (in limited number, without significant deviations~\cite{Wehle:2016yoi,Aaij:2015dea}) are compatible with no LFU-NP contributions (as often assumed), but they do not prevent these contributions from occurring. Assuming hadronic contributions properly assessed~\cite{Capdevila:2017ert,Arbey:2018ics}, we consider for the first time that LFU-NP contributions can exist for both $\C9$ and $\C{10}$, together with LFUV-NP contributions. It is important to remark that this is not the same as simply allowing for NP in electrons: we impose as a constraint in the fit that this contribution is the same for all leptons and work out the consequences of this identity. The key point to lift the degeneracy between the various contributions through the fit consists in considering together LFUV and LFD observables. The LFUV observables will constrain LFUV-NP contributions ($\C{i\ell}^V$), whereas LFD observables will be sensitive to the sum of LFUV-NP and LFU-NP contributions ($\C{i}^U+\C{i\ell}^V$). As we increase the number of parameters, we have more flexibility to describe the data, which could lead to an improvement compared to our earlier fits restricted to LFUV-NP contributions only and opens the possibility of new NP models.
 
We start from the results presented in the Table II of Ref.~\cite{Capdevila:2017bsm}, for the global fits under (favoured) 1D hypotheses of NP in $b\to s\mu\mu$. The 1D hypothesis with $\C{9\mu}^{\rm NP}$ (scenario 1) led to a 68\% confidence interval of  $[-1.28,-0.94]$ with a pull w.r.t. the SM of 5.8$\sigma$, whereas the hypothesis $\C{9\mu}^{\rm NP}=-\C{10\mu}^{\rm NP}$ (scenario 2)
had a 68\% confidence interval of $[-0.75,-0.49]$ with a pull of 5.3$\sigma$. We consider now a set of nested fits named scenarios 3 to 8 and presented in  Tables \ref{Fit4D}-\ref{FitC9} in decreasing order of complexity to better understand the interplay between LFUV and LFU NP (more information
and results, including the correlations among the parameters, are given in the Appendix).

\begin{table}[t]
\begin{tabular}{cccc}
\toprule 
 & Best-fit point & 1 $\sigma$ CI & 2 $\sigma$ CI \\
\hline
$\C{9\mu}^{\rm V}$ & $0.08$ & $[-0.72,0.80]$ & $[-1.69,1.49]$\\
$\C{10\mu}^{\rm V}$ & $1.14$ & $[0.66,1.59]$ & $[0.12,2.03]$\\
$\C{9}^{\rm U}$ & $-1.26$ & $[-1.92,-0.25]$ & $[-2.43,1.62]$\\
$\C{10}^{\rm U}$ & $-0.91$ & $[-1.40,-0.40]$ & $[-1.89,0.16]$\\
\hline
\end{tabular} 
\caption{Scenario 3: 4D hypothesis with $\C{9\mu}^{\rm V}$ and $\C{10\mu}^{\rm V}$, and with $\C{9}^{\rm U}$ and $\C{10}^{\rm U}$. Confidence Intervals (CI) are also provided.}\label{Fit4D}
\end{table}
\begin{table}[t]
\begin{tabular}{cccc}
\toprule 
 & Best-fit point & 1 $\sigma$ CI & 2 $\sigma$ CI \\
\hline
$\C{9\mu}^{\rm V}=-\C{10\mu}^{\rm V}$ & $-0.68$ & $[-0.96,-0.45]$ & $[-1.28,-0.26]$\\
$\C{9}^{\rm U}$ & $-0.37$ & $[-0.68,-0.03]$ & $[-0.95,0.35]$\\
$\C{10}^{\rm U}$ & $-0.51$ & $[-0.86,-0.18]$ & $[-1.24,0.13]$\\
\hline
\end{tabular} 
\caption{Scenario 4: 3D hypothesis with $\C{9\mu}^{\rm V}= - \C{10\mu}^{\rm V}$ and with $\C{9}^{\rm U}$ and $\C{10}^{\rm U}$.}\label{Fit3D}
\end{table}
\begin{table}[t]
\begin{tabular}{cccc}
\toprule 
 & Best-fit point & 1 $\sigma$ CI & 2 $\sigma$ CI \\
\hline
$\C{9\mu}^{\rm V}$ & $-0.16$ & $[-0.94,0.46]$ & $[-2.05,0.98]$\\
$\C{10\mu}^{\rm V}$ & $1.00$ & $[0.18,1.59]$ & $[-1.35,2.06]$\\
$\C{9}^{\rm U}=\C{10}^{\rm U}$ & $-0.87$ & $[-1.43,-0.14]$ & $[-1.91,0.98]$\\
\hline
\end{tabular} 
\caption{Scenario 5: 3D hypothesis with $\C{9\mu}^{\rm V}$ and $\C{10\mu}^{\rm V}$ and with $\C{9}^{\rm U} = \C{10}^{\rm U}$.}\label{Fit3Dbis}
\end{table}

\begin{table}[t]
\begin{tabular}{cccc}
\toprule 
 & Best-fit point & 1 $\sigma$ CI & 2 $\sigma$ CI \\
\hline
$\C{9\mu}^{\rm V}=-\C{10\mu}^{\rm V}$ & $-0.64$ & $[-0.77,-0.51]$ & $[-0.90,-0.39]$\\
$\C{9}^{\rm U}=\C{10}^{\rm U}$ & $-0.44$ & $[-0.58,-0.29]$ & $[-0.71,-0.14]$\\
\hline
\end{tabular} 
\caption{Scenario 6: 2D hypothesis with $\C{9\mu}^{\rm V} = - \C{10\mu}^{\rm V}$ and $\C{9}^{\rm U} = \C{10}^{\rm U}$.}\label{Fit2D}
\end{table}

\begin{table}[t]
\begin{tabular}{cccc}
\toprule 
 & Best-fit point & 1 $\sigma$ CI & 2 $\sigma$ CI \\
\hline
$\C{9\mu}^{\rm V}$ & $-1.57$ & $[-2.14,-1.06]$ & $[-2.75,-0.58]$\\
$\C{9}^{\rm U}$ & $0.56$ & $[0.01,1.15]$ & $[-0.51,1.78]$\\
\hline
$\C{9\mu}^{\rm V}=-\C{10\mu}^{\rm V}$ & $-0.42$ & $[-0.57,-0.27]$ & $[-0.72,-0.15]$\\
$\C{9}^{\rm U}$ & $-0.67$ & $[-0.90,-0.42]$ & $[-1.11,-0.16]$\\
\hline
\end{tabular} 
\caption{2D hypotheses. Top: Scenario 7: LFUV and LFU NP in $\C{9\ell}^{\rm NP}$ only. Bottom: Scenario 8: $\C{9\mu}^{\rm V}= - \C{10\mu}^{\rm V}$ and $\C{9}^{\rm U}$ only.}\label{FitC9}
\end{table}

\begin{itemize}

\item The general hypothesis $\{ \C{9\mu}^{\rm V} , \C{10\mu}^{\rm V},\C{9}^{\rm U} , \C{10}^{\rm U} \}$ (Table~\ref{Fit4D}) has a pull of 5.6$\sigma$ w.r.t. the SM. The result is remarkable: considering the best-fit point (b.f.p.), $\C{9\mu}^{\rm V}$ almost vanishes, $\C{9,10}^{\rm U}$ are far away from zero, and $\C{10\mu}^{\rm V}$ is larger than $1$. 
At first glance, this result seems to contradict the previous global analyses (including Ref.~\cite{{Capdevila:2017bsm}}) and should be explained in more detail.
The key observation is that $R_{K^{(*)}}$-like observables may be also accommodated by $\C{10\mu}^{\rm V}$ alone with a negligible $ \C{9\mu}^{\rm V}$, cf. Appendix. This result was not obtained in the 2D fits with only LFUV-NP contributions (setting $\C{i}^U=0$), since LFD observables led then to the favoured scenarios with b.f.p. $ \C{9\mu}^{\rm V}\simeq -1$ and $ \C{9\mu}^{\rm V} = - \C{10\mu}^{\rm V} \simeq -0.7$. 
Adding LFU contributions provides complementary mechanisms to explain LFUV and LFD anomalies. On one side, the LFD anomalies are accommodated by $ \C{9\mu}^{\rm V} + \C9^{\rm U} \simeq -1.18$ and  $ \C{10\mu}^{\rm V} + \C{10}^{\rm U} \simeq +0.23$. On the other side, the LFUV observables are accommodated by $\C{10\mu}^{\rm V} \simeq 1.14 $. It is thus not a surprise that the summed LFU and LFUV contributions for both $\C{9,10}$ yield a result close to the fit to all observables under the NP hypothesis $\left( \C{9\mu}^{\rm NP}, \C{10\mu}^{\rm NP} \right)$ showed in Table III of Ref.~\cite{Capdevila:2017bsm}.
Under this hypothesis, $\C{9\mu}^{\rm V}$ changes sign w.r.t. fits without LFU in order to resolve the inner tensions between LFUV and LFD observables. Moreover the constraint from ${\cal B}({B_s \to \mu^+ \mu^-})$  is obeyed by the sum $ \C{10\mu}^{\rm V} + \C{10}^{\rm U}$ with opposite signs and thus allowing a large $ \C{10\mu}^{\rm V}$. This important feature is observed for the first time here and opens new possibilities for models beyond the SM.

\item The hypothesis $\{ \C{9\mu}^{\rm V}=-\C{10\mu}^{\rm V},\C{9}^{\rm U} , \C{10}^{\rm U} \}$ (Table~\ref{Fit3D}) is model building motivated for theories with a significant scale gap between SM and NP~\cite{Buttazzo:2017ixm,Buchmuller:1985jz,Grzadkowski:2010es,Alonso:2014csa,Celis:2017doq}, as the additional NP contributions should be invariant under $SU(2)_L$. 
Remarkably, this 3D fit has a pull of 5.7$\sigma$ to SM. The b.f.p. is in good agreement with the result found in Table II of Ref.~\cite{Capdevila:2017bsm} but with LFU contributions differing from zero at the 1$\sigma$ level. The increase in the SM pull w.r.t. the case without LFU (5.7$\sigma$ versus 5.3$\sigma$~\cite{Capdevila:2017bsm}, with 2 more parameters) hints at a slight preference for LFU-NP in both $\C{9,10}$.

\item The hypothesis $\{ \C{9\mu}^{\rm V}, \C{10\mu}^{\rm V},\C{9}^{\rm U} = \C{10}^{\rm U} \}$ (Table~\ref{Fit3Dbis}) is inspired by the fit in Table~\ref{Fit4D} which suggests $\C{9}^{\rm U} \simeq \C{10}^{\rm U}$. We find  Table~\ref{Fit3Dbis} with a pull of 5.8$\sigma$ w.r.t. SM, slightly larger than the 4D hypothesis.

\item The hypothesis $\{ \C{9\mu}^{\rm V} = -\C{10\mu}^{\rm V},\C{9}^{\rm U} = \C{10}^{\rm U} \}$ (Table~\ref{Fit2D}) combines the suggestive results  from both Table~\ref{Fit4D} and Table~\ref{Fit3Dbis} and yields a fit with a pull of 6.0$\sigma$ w.r.t. the SM. The b.f.p. $\C{9\mu}^{\rm V} = -\C{10\mu}^{\rm V} = -0.64$ obtained now is very similar to the one from Table~\ref{Fit3D}, and $\C{9}^{\rm U} = \C{10}^{\rm U} = -0.44$ is exactly the average of LFU contributions found in Table~\ref{Fit3D}. This particular 2D correlation is shown in Fig.~\ref{fig:correlscenario6} (see the Appendix for correlations under the other hypotheses). It is interesting that a $\C{10}$ contribution gives rise to a rather tight 1$\sigma$ confidence interval, mainly due to ${\cal B}({B_s \to\mu^+\mu^-})$. Let us add that the hypothesis  $\{ \C{9\mu}^{\rm V} = -\C{10\mu}^{\rm V},\C{9}^{\rm U} =-\C{10}^{\rm U} \}$ (once again of interest for models based on $SU(2)_L$ invariance) has a pull w.r.t. the SM lower by almost $1 \sigma$ and is not favoured by the data, as can already be seen in Table~\ref{Fit4D}.
\end{itemize}

Two additional 2D hypotheses provide a bridge between the above hypotheses with LFU NP  in both $\C{9,10}$ and previous results focused on LFUV-NP contributions in $\C{9\mu}^{\rm NP}$ without LFU-NP. We consider two 2D fits:
\begin{itemize}

\item The hypothesis $\{ \C{9\mu}^{\rm V}, \C9^{\rm U} \}$ (Table~\ref{FitC9}, top) has  a pull w.r.t. the SM of 5.7 $\sigma$. The LFD observables are governed by the sum $\C{9\mu}^{\rm V} + \C{9}^{\rm U}\simeq -1.01$ for the b.f.p.,  whereas the b.f.p. $\C{9\mu}^{\rm V} = -1.57$ is the dominant contribution to LFUV observables. Interestingly
these results can be linked to the results of Ref.~\cite{Capdevila:2017bsm} (without LFU NP) 
with the former  in agreement with the b.f.p. of the fit to all data ($-1.11$) in Table II of \cite{Capdevila:2017bsm} and the latter closer to the b.f.p. of the fit to LFUV observables ($-1.76$). Therefore the internal tension between LFD and LFUV observables in the global fit of Ref.~\cite{Capdevila:2017bsm} is resolved here due to the additional freedom allowed by $\C{9\mu}^{\rm V}$, which enters the LFUV observables (always with a subleading contribution from $\C{9\mu}^{\rm U}$) whereas the combination  $\C{9\mu}^{\rm V} + \C{9}^{\rm U}$ is constrained by the LFD observables.

\item The hypothesis $\{ \C{9\mu}^{\rm V} = - \C{10\mu}^{\rm V}, \C9^{\rm U} \}$ (Table~\ref{FitC9}, bottom) has a  pull of 5.8$\sigma$ and follows a similar pattern. The LFUV contribution $\C{9\mu}^{\rm V}= - \C{10\mu}^{\rm V} = -0.42$ (for the b.f.p.) accommodates well the LFUV observables, while the sum $\C{9\mu}^{\rm V} + \C{9}^{\rm U}  = - 1.09$ takes care of the LFD observables (recovering approximately the fits to all data and to LFUV observables only from Ref. \cite{Capdevila:2017bsm}).

\end{itemize}

The similar pulls w.r.t. the SM of the various scenarios indicate that the
current measurements cannot lift the degeneracy among the hypotheses, and a different strategy should be envisaged in order to distinguish them. 

\begin{figure}[t]
\begin{center}
\includegraphics[width=8cm]{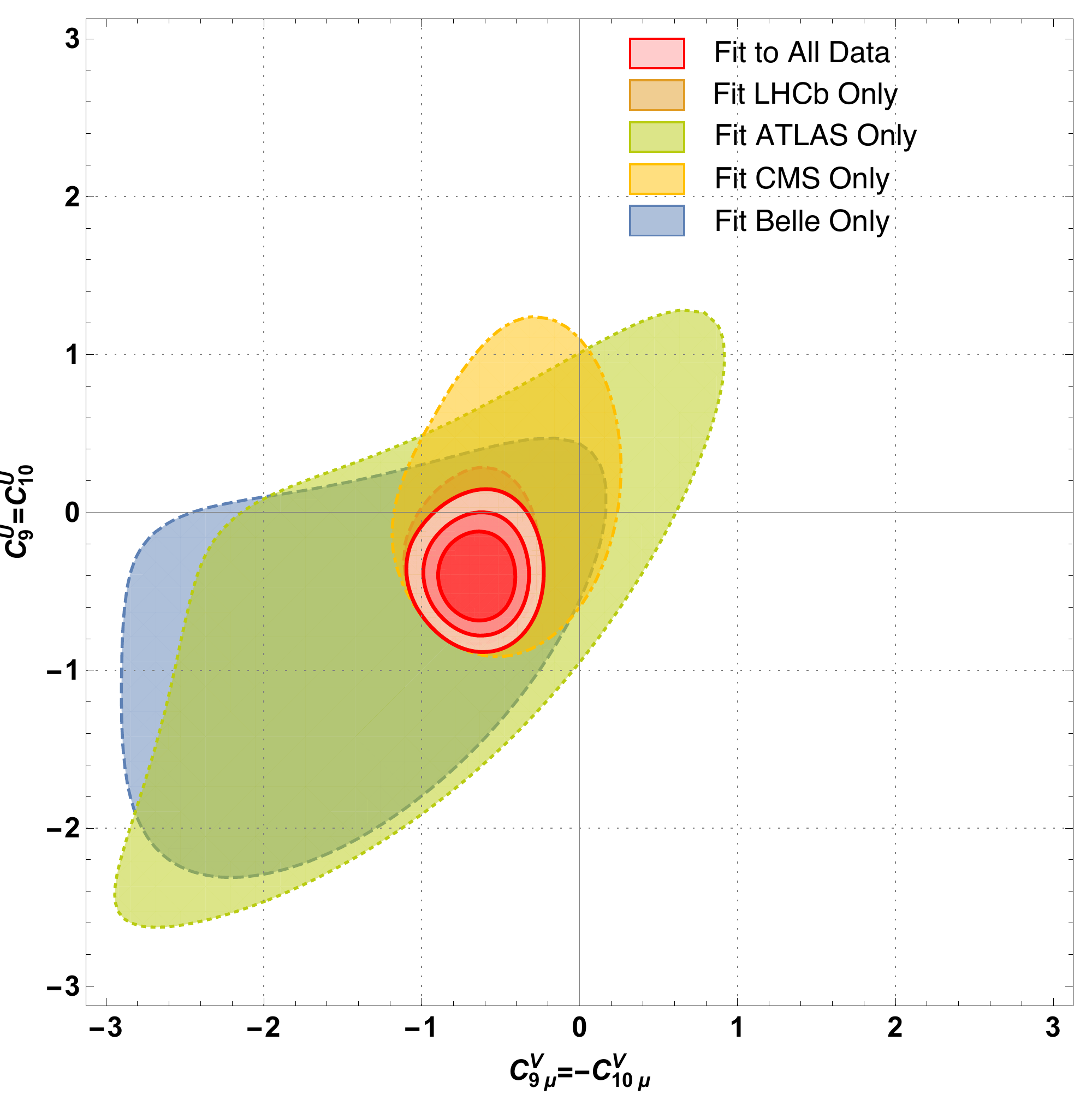}
\caption{Confidence regions for scenario 6 in the plane $(\C{9\mu}^{\rm V} = - \C{10\mu}^{\rm V},\C{9}^{\rm U} = \C{10}^{\rm U})$. The regions for different experimental subsets correspond to a confidence level of 3$\sigma$, wheres the 1,2,3$\sigma$ confidence regions are shown for the region associated with the global fit to all data.}
\label{fig:correlscenario6}
\end{center}
\end{figure}

\begin{figure}[t]
\begin{center}
\includegraphics[width=8cm]{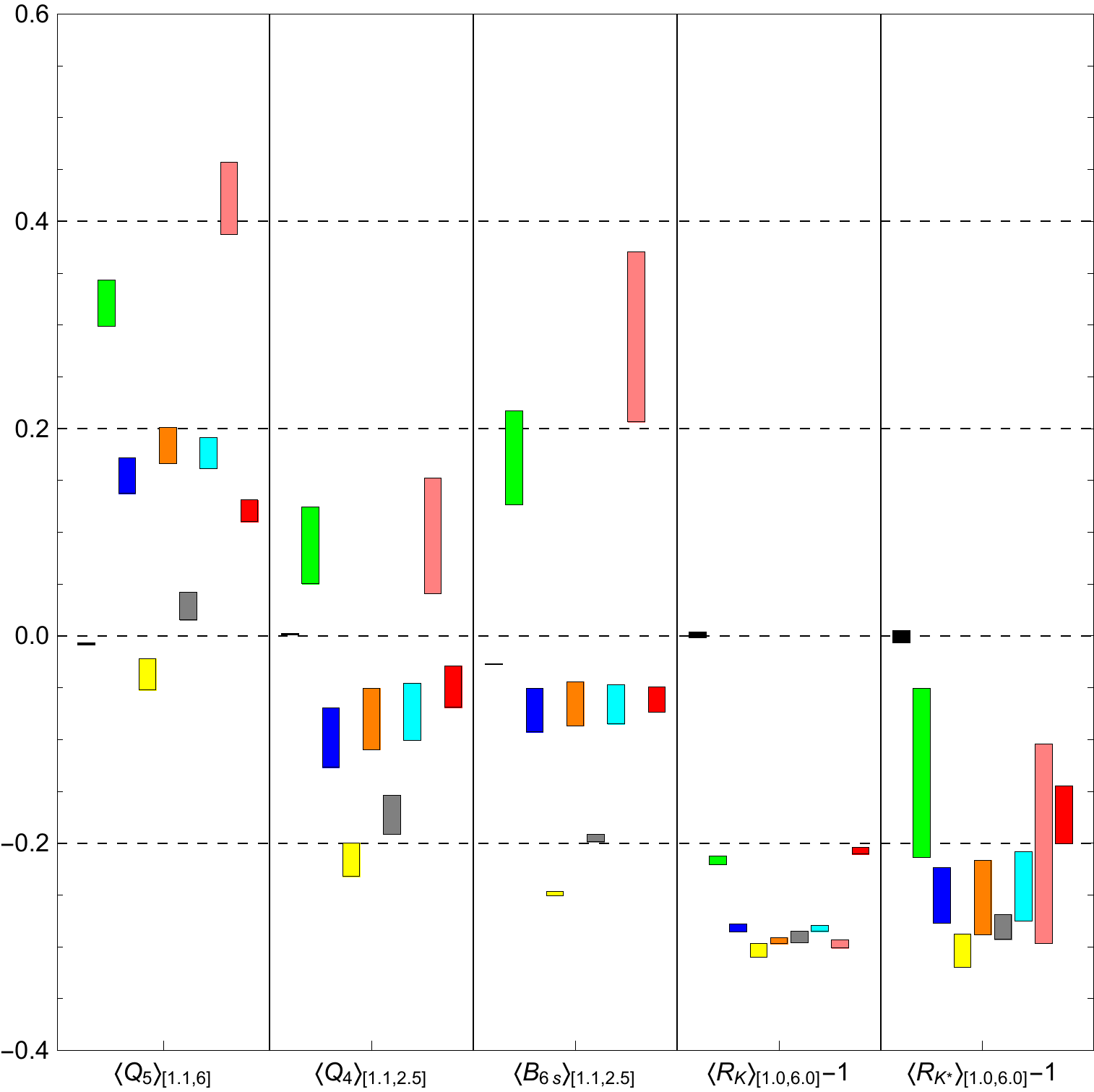}
\caption{Predictions for LFUV observables of interest under various hypotheses of LFU and LFUV-NP contributions currently favoured by the global $b\to s\ell\ell$. From left to right: SM is followed by Scenarios 1 to 8 as described in the main text. We plot $R_{K,K^*}-1$ to keep the figure compact.}
\label{fig:plotpreds}
\end{center}
\end{figure}

\section{The role of LFUV observables}

One of the most relevant outcomes of this work is the unexpected preference for a NP solution with a prominent $\C{10\ell}^{\rm NP}$ signature, both LFUV and LFU. This may represent a shift of paradigm, since until now the vast majority of global analyses  performed were signalling a single NP contribution to $\C{9\mu}^{\rm V}$ as the most favoured solution.

The LFUV observables are natural candidates in order to identify the contributions from LFUV-NP conclusively. While new and more precise measurements of $R_K$ and $R_{K^*}$ will be certainly useful, Refs.~\cite{Capdevila:2016ivx, Capdevila:2017bsm} pointed out the relevance of the $Q_i$ observables (difference of optimised angular observables in muon and electron modes) and the more exotic $B_{5,6s}$ observables.
Indeed these observables are not only very clean and stringent tests against the SM, similarly to $R_{K^{(*)}}$, but they also contain additional information on the Wilson coefficients from a full angular analysis. In particular, while $R_K$ involves crossed LFUV-LFU terms such as $\C{9 \mu}^{\rm V} \C{10}^{\rm U}$ and $\C{10 \mu}^{\rm V} \C{9}^{\rm U}$, $Q_{5}$ contains a $\C{9 \mu}^{\rm V} \C{9}^{\rm U}$ term, introducing complementary information to $R_K$, see Appendix. 

A natural candidate to disentangle LFU and LFUV NP is then $\langle Q_5 \rangle_{[1.1,6]}$ because of its high sensitivity to $\C{9\mu}^{\rm V}$ and its ties to the $P'_5$ anomaly sensitive to both types of NP contributions. $\langle R_K\rangle_{[1,6]}$ and in second place $\langle R_{K^*}\rangle_{[1,6]}$, despite the large theoretical uncertainties of the latter in presence of NP, should also play a role due to their sensitivity to $\C{10\mu}^{\rm V}$. Finally, the very same $\langle P_5^{'} \rangle_{[4,6]}$ should help to discern between the LFU contributions $\C{9}^{\rm U}$ and $\C{10}^{\rm U}$. 

We show the most interesting LFUV observables for the b.f.p. of the above scenarios in Fig.~\ref{fig:plotpreds} 
(from left to right, SM and Scenarios 1 to 8). Explicit expressions of these observables are given in the Appendix.

A decision tree can be built from the experimental measurement of $\langle Q_5 \rangle_{[1.1,6]}$ which exhibits a good discriminating power against the various scenarios considered above:

\begin{itemize}
\item If $\langle Q_5\rangle_{[1.1,6]}\gtrsim 0.3$ (first column in Fig.~\ref{fig:plotpreds}), the 1D hypothesis $\C{9\mu}^{\rm V}$ is able to explain all anomalies. A confirmation can come from an updated measurement of $\langle R_{K^*}\rangle_{[1,6]} -1\gtrsim -0.2$ (last column in Fig.~\ref{fig:plotpreds}).

\item If $0.1\lesssim\langle Q_5\rangle_{[1.1,6]}\lesssim 0.2$, the hypotheses with only a large $\C{9\mu}^{\rm NP}$ are disfavoured while hypotheses with $\C{10\mu}^{\rm V}$ are favoured. Actually, this range of values corresponds to solutions involving $\C{9\mu}^{\rm V}=-\C{10\mu}^{\rm V}$ (scenarios 2, 4, 6, 8). Knowing $\langle R_K\rangle_{[1,6]}$ with an uncertainty around $5\%$ would help discriminating between the hypotheses $\C{9\mu}^{\rm V}=-\C{10\mu}^{\rm V}$ with and without LFU contributions in $\C{9\mu}^{\rm NP}$ (scenarios 2 and 8 respectively, see fourth column in Fig.~\ref{fig:plotpreds}). $\langle P'_5 \rangle_{[4,6]}$ can confirm this result by disentangling $\{ \C{9}^{\rm U} , \C{10\mu}^{\rm U} \}$ from $\{ \C{9}^{\rm U} =\C{10\mu}^{\rm U} \}$, see Appendix, if the experimental uncertainty on $\langle P'_5 \rangle_{[4,6]}$ is reduced by half~\cite{Descotes-Genon:2015uva}.

\item If $\langle Q_5\rangle_{[1.1,6]}\lesssim0.1$, scenarios where $\C{9\mu}^{\rm V}$, $\C{10\mu}^{\rm V}$  are left free to vary independently (scenarios 3, 5) are preferred. Distinguishing among these two scenarios is practically impossible, since only $\langle B_{6s}\rangle_{[1.1,2.5]}$ shows a very mild discrimination power (third column in Fig.~\ref{fig:plotpreds})
if measured at a very high  precision.
\end{itemize}

The value of $\C{10}^{\rm U}$ can be probed by ${\cal B}(B_s\to \ell^+\ell^-)$, assuming no significant scalar or pseudoscalar contributions:
\begin{equation}
\frac{{\cal B}({B_s \to e^+ e^-})}{{\cal B}({B_s \to \mu^+ \mu^-})} = \frac{m_e^2}{m_\mu^2} \times \frac{|\C{10}^{\rm SM}+\C{10}^{\rm U}|^2}{|\C{10}^{\rm SM}+\C{10\mu}^{\rm V}+\C{10}^{\rm U}|^2}.
\end{equation}
The inclusion of $\C{10}^{\rm U}$ in this equation leads to an enhancement between 30-60$\%$ w.r.t. the SM prediction, but with strong lepton-mass suppression for this observable to be available in the near future, and similarly, assuming no large LFUV-NP contributions in $b\to s\tau\tau$, for the challenging measurement of ${\cal B}({B_s \to \tau^+ \tau^-})$.

\section{Conclusions}

We have considered the current anomalies observed in $b\to s\ell\ell$ transitions and discussed the consequences of removing one hypothesis frequently made (and overlooked) in the global model-independent analyses, namely that the anomalies are explained only by NP violating Lepton Flavour Universality. Instead we explore the implications of allowing both LFU and LFUV-NP contributions in the Wilson coefficients $\C{9}$ and $\C{10}$, providing more flexibility to describe the data.  The LFUV observables will constrain LFUV-NP contributions, whereas LFD observables will be sensitive to the sum of LFUV-NP and LFU-NP contributions. We found a different mechanism with a large contribution to $\C{10}$ to explain the data without transgressing the ${\cal B}({B_s \to \mu^+\mu^-})$ constraint, leading to an improvement compared to our earlier fits restricted to LFUV-NP contributions only.

The 4D hypothesis with both kind of contributions to $\C{9\ell}^{\rm NP}$ and $\C{10\ell}^{\rm NP}$ leads to two  
scenarios with high significances and well-constrained parameters (equivalent to scenarios with only LFUV-NP contributions and thus a more limited set of parameters). Indeed, the fits favour either a large and positive $ \C{10\mu}^{\rm V}$ together with large and negative LFU contributions in both $\C{9,10}^{\rm U}$ (scenarios 3 and 5, Tables~\ref{Fit4D}, \ref{Fit3Dbis}), or a  negative $\C{9\mu}^{\rm V} = - \C{10\mu}^{\rm V}$ together with smaller (in absolute value) but still negative LFU contributions in both $\C{9,10}^{\rm U}$ (scenarios 4 and 6, Tables~\ref{Fit3D}, \ref{Fit2D}). If LFUV lepton interactions with V-A are favoured suggesting that $SU(2)_L$ invariance might be a guide for models for NP in $b\to s\ell\ell$, LFU lepton interactions with a V+A structure are preferred. The size and structure of these LFU lepton interactions do not agree with a generation by radiative effects from LFUV-NP contributions, which would lead to much smaller and purely vector LFU lepton interactions~\cite{Crivellin:2018yvo}. The scenarios that we discuss would also require a deviation from popular model-building ideas relying on a strong hierarchy of NP contributions according to the generations involved in order to provide a connection with $b\to c\tau\nu$ anomalies~\cite{Bifani:2018zmi,Capdevila:2017iqn,Glashow:2014iga,Buttazzo:2017ixm,Bhattacharya:2014wla,Alonso:2015sja,Calibbi:2015kma,Fajfer:2015ycq,Bauer:2015knc,Becirevic:2016yqi}.

To separate the various scenarios explaining the $b\to s\ell\ell$ anomalies, a decision tree is proposed. Although the update of $R_K$ will be a major milestone, the measurement of $Q_5$ (and the improvement of $R_{K^*}$ and $P'_5$) remains essential to disentangle the possible scenarios of NP and to interpret the effective description in terms of a full-fledged UV-complete model of physics beyond the Standard Model.

\section{Acknowledgments}

This work received financial support from European Regional Development Funds under the Spanish Ministry of Science, Innovation and Universities (projects No. FPA2014-55613-P and No. FPA2017-86989-P) and from the Agency for Management of University and Research Grants of the Government of Catalonia (project SGR 1069) [M. A., B. C., P.M., J. M.]; and from European Commission (Grant Agreements No. 690575, No. 674896 and No. 69219) [S. D. G.]. The work of P. M. is supported by the Beatriu de Pin—s postdoctoral program co-funded by the Agency for Management of University and Research Grants of the Government of Catalonia and by the COFUND program of the Marie Sklodowska-Curie actions under the framework program Horizon 2020 of the European Commission. J.M. gratefully acknowledges the financial support by ICREA under the ICREA Academia programme.


\appendix*

\section{}

\subsection{Polynomial parametrisation for some observables of interest}\label{sec:expressions}

The observables $\langle P'_5\rangle_{[4,6]}$, $\langle Q_5\rangle_{[1.1,6]}$, $\langle R_K\rangle_{[1,6]}$ and $\langle R_{K^*}\rangle_{[1,6]}$ can be parameterised as follows, with the coefficients $\alpha_i$ for each observable collected in Table~\ref{expressions}:

\begin{align}\nonumber
O_i&=\alpha_0+\alpha_1\ \C{9}^{\rm U}+\alpha_2\ \C{10}^{\rm U}+\alpha_3\ \C{9\mu}^{\rm V}+\alpha_4\ \C{10\mu}^{\rm V}+\alpha_5\ \left(\C{9}^{\rm U}\right)^2\\ \nonumber
&+\alpha_6\ \left(\C{10}^{\rm U}\right)^2+\alpha_7\ \left(\C{9\mu}^{\rm V}\right)^2+\alpha_8\ \left(\C{10\mu}^{\rm V}\right)^2 +\alpha_9\ \C{9}^{\rm U}\C{10}^{\rm U}\\ \nonumber
&+\alpha_{10}\ \C{9}^{\rm U}\C{9\mu}^{\rm V}+\alpha_{11}\ \C{9}^{\rm U}\C{10\mu}^{\rm V}+\alpha_{12}\ \C{9\mu}^{\rm V}\C{10}^{\rm U}+\alpha_{13}\ \C{9\mu}^{\rm V}\C{10\mu}^{\rm V} \\ 
& +\alpha_{14}\ \C{10}^{\rm U}\C{10\mu}^{\rm V} \label{eq:expressions}
\end{align}

\vspace{3mm}

\begin{widetext}
\begin{center}
\begin{table}[h]
\begin{tabular}{c||c||c|c|c|c||c|c|c|c|c|c|c|c}
& $\alpha_0$ & $\alpha_{1}$ & $\alpha_{2}$ & $\alpha_{3}$ & $\alpha_{4}$ & $\alpha_{5}$ & $\alpha_{7}$ &  $\alpha_{9}$ & $\alpha_{10}$ & $\alpha_{11}$ & $\alpha_{12}$ & $\alpha_{13}$ & $\alpha_{14}$ \\
\hline \hline
$\langle P'_5\rangle_{[4,6]}$ & $-0.814$ & $-0.207$ & $-0.066$ & $-0.207$ & $-0.066$ & $0.058$ &  $0.058$ & $0.011$ & $0.116$ & $0.011$ & $0.011$ & $0.011$ & $0.008$ \\
\hline
$\langle Q_5\rangle_{[1.1,6]}$ & $0$ & $0$ & $0$ & $-0.246$ & $-0.019$ & $0$ & $0$ & $0.033$  & $0.066$ & $0$ & $0$ & $0$ & $0.013$ \\
\hline
$\langle R_K\rangle_{[1,6]}$ & $1.001$ & $0$ & $0$ & $0.230$ & $-0.264$ & $0$ & $0$ & $0$ & $0$ & $0.061$ & $0.061$ & $0$ & $0$ \\
\hline
$\langle R_{K^*}\rangle_{[1,6]}$ & $1.000$ & $0$ & $0$ & $0.157$ & $-0.287$ & $0$ & $0$ & $0$ & $0.042$ & $0.045$ & $0.045$ & $0$ & $-0.016$ \\
\end{tabular}
\caption{Coefficients of the polynomial parametrisation of observables in Eq.~(\ref{eq:expressions}).} \label{expressions}
\end{table} 
\end{center}
\end{widetext}

The first block in Table~\ref{expressions} (second column) contains the Standard Model prediction. In the second block (columns three to six) one can find the coefficients of the linear terms ($\C{9}^{\rm U},\C{10}^{\rm U},\C{9\mu}^{\rm V},\C{10\mu}^{\rm V}$) and the third block shows the coefficients of the quadratic terms. Since in the four observables the terms $\alpha_{6,8}$ are zero we have not included them in Table~\ref{expressions}. 

$\langle P'_5\rangle_{[4,6]}$ being the only LFD observable in Table~\ref{expressions} is obviously  the only observable with non-zero linear LFU terms. The combination $\C{9}^{\rm U}+\C{9\mu}^{\rm V}$ dominates the expression being $\alpha_1=\alpha_3=-0.207$, which is $\sim 25\%$ of the SM value. The coefficients in front of  $\C{10,\mu}^{\rm U,V}$ verify $\alpha_{2,4}\sim1/3\alpha_{2,3}$ while the coefficients of the quadratic terms $\left(\C{9,\mu}^{\rm U,V}\right)^2$ are $\alpha_{5,7}\sim\alpha_{2,4}$. Moreover, $\langle P'_5\rangle_{[4,6]}$ also has crossed terms mixing $\C{i}^{\rm U}$ and $\C{j\mu}^{\rm V}$ ($\alpha_{9...14}$), even though they are subleading with respect to $\alpha_{1,3}$. 

$\langle Q_5\rangle_{[1.1,6]}$ is strongly sensitive to $\C{9\mu}^{\rm V}$, with $\alpha_3=-0.246$ being an order of magnitude larger than the rest of the coefficients. $\langle R_K\rangle_{[1,6]}$ and  $\langle R_{K^*}\rangle_{[1,6]}$ are linearly sensitive to both $\C{9,10\mu}^{\rm V}$ but the former only contains crossed terms mixing universal and violating contributions of the type $\C{9(10)}^{\rm U}\C{10(9)\mu}^{\rm V}$. Contrarily, the latter $\langle R_{K^*}\rangle_{[1,6]}$ has also quadratic terms such as $\C{9(10)}^{\rm U}\C{9(10)\mu}^{\rm V}$. This implies that if one sets either $\C{9}$ or $\C{10}$ to zero, this kind of terms remain in $\langle R_{K^*}\rangle_{[1,6]}$ while they vanish in $\langle R_{K}\rangle_{[1,6]}$. This difference in structure can prove useful in disentangling different scenarios.

The coefficients $\alpha_i$ of these parameterisations have been obtained by fitting the calculated expressions of the observables with the second-order polynomial in the Wilson coefficients in Eq.~(\ref{eq:expressions}). We generated the central values over a grid of values of $\C{9\mu}^{\rm NP},\C{10\mu}^{\rm NP},\C{9e}^{\rm NP},\C{10e}^{\rm NP}$. The grid range of the grid varied from [-1,1] for the coefficients $\C{10\mu}^{\rm NP},\C{9e}^{\rm NP},\C{10e}^{\rm NP}$ to [-2,2] for the coefficient $\C{9\mu}^{\rm NP}$ (with a spacing of the grid of sampled points of 0.1). We stress that the above formulae correspond to central values only (the associated uncertainties could be parameterised in a similar way), but they already help in identifying the main sensitivities of these observables.

\begin{figure*}[hbt]
\centering
\includegraphics[width=0.42\linewidth]{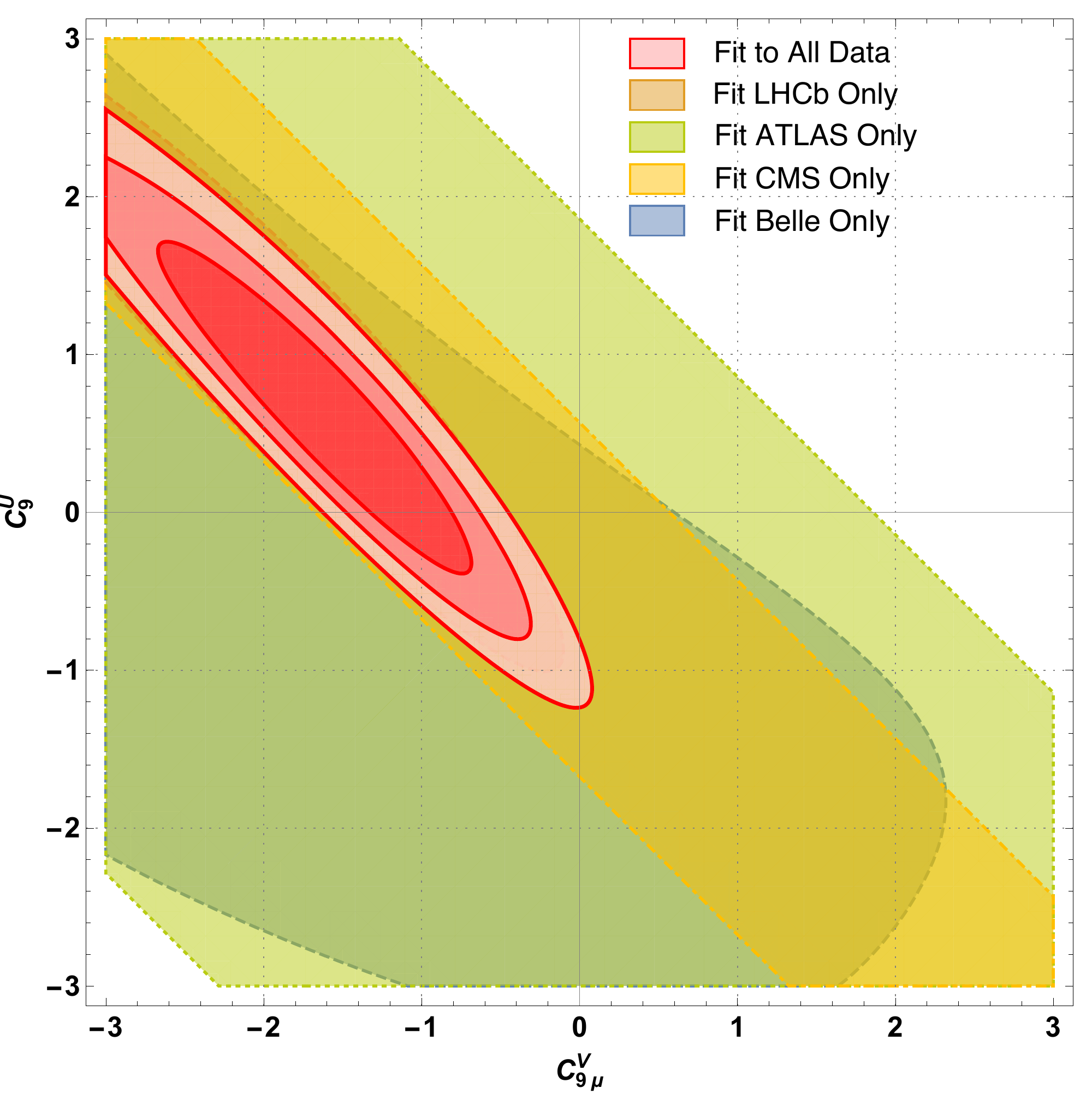}
\includegraphics[width=0.42\linewidth]{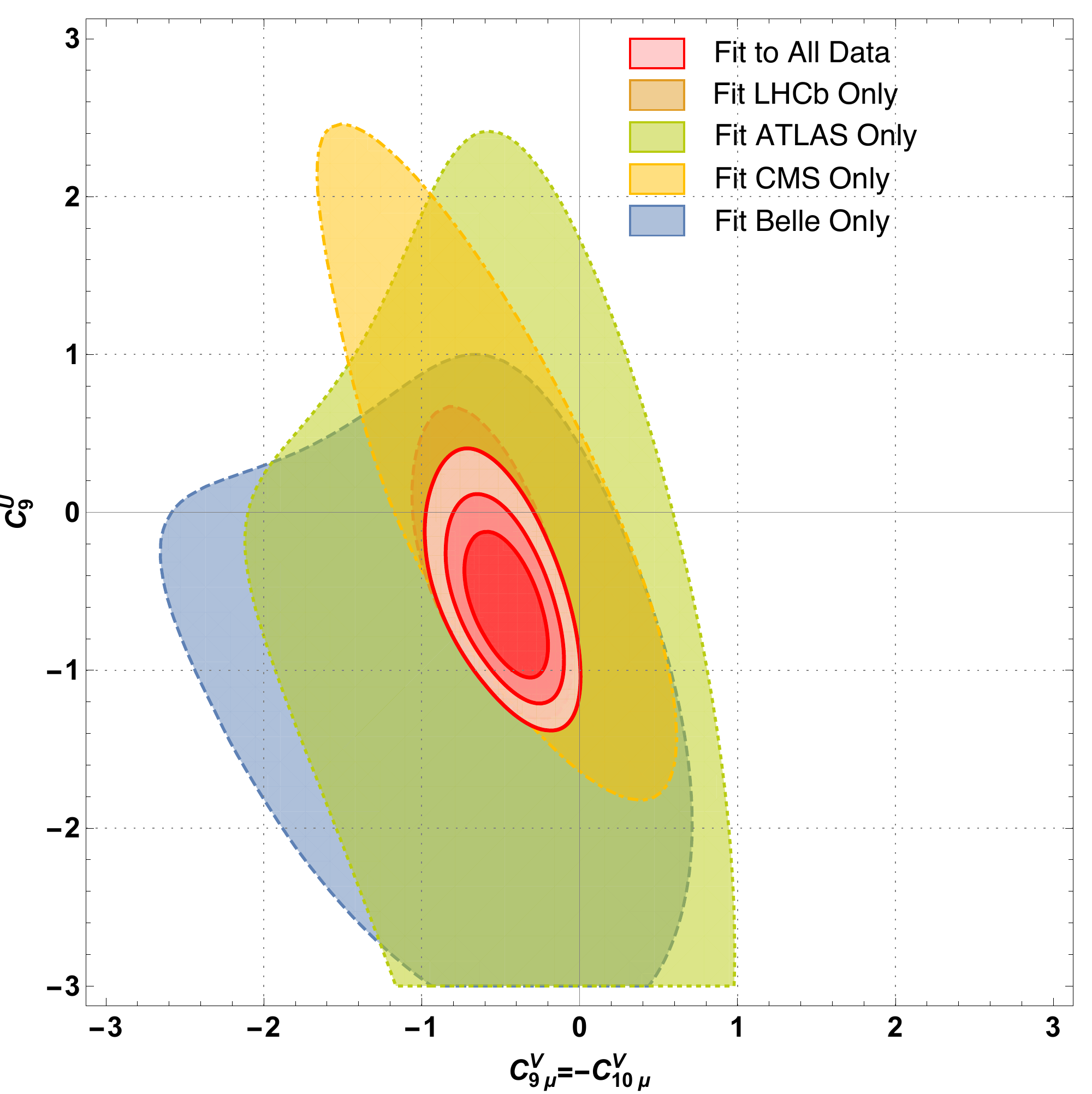}
\caption{{\bf Left:} Correlation between $\C{9\mu}^{\rm V}$ and $\C{9}^{\rm U}$ from the Scenario 7 fit. {\bf Right:} Correlation between the parameters $\C{9\mu}^{\rm V}=-\C{10\mu}^{\rm V}$ and $\C{9}^{\rm U}$ from the Scenario 8 fit. The regions for different experimental subsets correspond to a confidence level of 3$\sigma$, wheres the 1,2,3$\sigma$ confidence regions are shown for the region associated with the global fit to all data.}
\label{fig:plots2D}
\end{figure*}

\subsection{Correlations among parameters of the fits}\label{sec:correlations}

Figure~\ref{fig:plots2D} shows the 1, 2 and $3\sigma$ confidence regions of the 2D fits. We also provide information about the correlations between the different parameters of each of the fits performed.

The correlations between the parameters of each fit are the following (in the order of the parameters given to describe each scenario):

\begin{itemize}
\item Scenario 3 - $\{\C{9\mu}^{\rm V},\C{10\mu}^{\rm V},\C{9}^{\rm U},\C{10}^{\rm U}\}$:\begin{equation*} 
{\rm Corr_3}=\left(\begin{array}{ccccc} 
1.00 & 0.59 & -0.96 & -0.52 &  \\
0.59 & 1.00 & -0.56 & -0.91 &  \\
-0.96 & -0.56 & 1.00 & 0.48 &  \\
-0.52 & -0.91 & 0.48 & 1.00 &  \\
\end{array} \right) 
 \end{equation*} 
 \item Scenario 4 - $\{\C{9\mu}^{\rm V}=-\C{10\mu}^{\rm V},\C{9}^{\rm U},\C{10}^{\rm U}\}$: \begin{equation*} 
{\rm Corr_4}=\left(\begin{array}{cccc} 
1.00 & -0.76 & 0.77 &  \\
-0.76 & 1.00 & -0.64 &  \\
0.77 & -0.64 & 1.00 &  \\
\end{array} \right) 
 \end{equation*} 
 \item Scenario 5 - $\{\C{9\mu}^{\rm V},\C{10\mu}^{\rm V},\C{9}^{\rm U}=\C{10}^{\rm U}\}$: \begin{equation*} 
{\rm Corr_{5}}=\left(\begin{array}{cccc} 
1.00 & 0.88 & -0.93 &  \\
0.88 & 1.00 & -0.92 &  \\
-0.93 & -0.92 & 1.00 &  \\
\end{array} \right) 
\end{equation*} 
\item Scenario 6 - $\{\C{9\mu}^{\rm V}=-\C{10\mu}^{\rm V},\C{9}^{\rm U}=\C{10}^{\rm U}\}$: \begin{equation*} 
{\rm Corr_6}=\left(\begin{array}{ccc} 
1.00 & -0.01 &  \\
-0.01 & 1.00 &  \\
\end{array} \right) 
 \end{equation*} 
 \item Scenario 7 - $\{\C{9\mu}^{\rm V},\C{9}^{\rm U}\}$: \begin{equation*} 
{\rm Corr_7}=\left(\begin{array}{ccc} 
1.00 & -0.93 &  \\
-0.93 & 1.00 &  \\
\end{array} \right) 
 \end{equation*} 
 \item Scenario 8 - $\{\C{9\mu}^{\rm V}=-\C{10\mu}^{\rm V},\C{9}^{\rm U}\}$: \begin{equation*} 
{\rm Corr_8}=\left(\begin{array}{ccc} 
1.00 & -0.47 &  \\
-0.47 & 1.00 &  \\
\end{array} \right) 
 \end{equation*} 
\end{itemize}

The 4D fit (Scenario 3) exhibit very strong anticorrelations between $\C{9\mu}^{\rm V}$ and $\C{9}^{\rm U}$, and between $\C{10\mu}^{\rm V}$ and $\C{10}^{\rm U}$. This is logical since $b\to s\mu^{+}\mu^{-}$ constrains $\C{i\mu}^{\rm V}+\C{i}^{\rm U}$ while $b\to se^{+}e^{-}$ constrains $\C{i}^{\rm U}$ and the LFUV observables constrain $\C{i\mu}^{\rm V}$. In fact, without LFUV observables we would find a correlation of -1 between $\C{i\mu}^{\rm V}$ and $\C{i}^{\rm U}$ because the LFD observables only see the sum of both types of contributions. The same pattern can be observed in the other fits, although correlations are nominally less strong due to the fact that different and more involved structures, like $\C{9\mu}^{\rm V}=-\C{10\mu}^{\rm V}$, are explored.

One should also stress that the correlation between the parameters $\C{9\mu}^{\rm V}=-\C{10\mu}^{\rm V}$ and $\C{9}^{\rm U}=\C{10}^{\rm U}$ in Scenario 6 is negligible, signaling at its statistical independence. This means that the underlying structure of most of the LFD observables is such that, when imposing $\C{9\mu}^{\rm V}=-\C{10\mu}^{\rm V}$, once its value is fitted to the LFUV observables, the parameter $\C{9}^{\rm U}=\C{10}^{\rm U}$ can be independently determined by the LFD observables.

\subsection{Further tests}
\begin{figure}[h]
\includegraphics[width=0.45\textwidth]{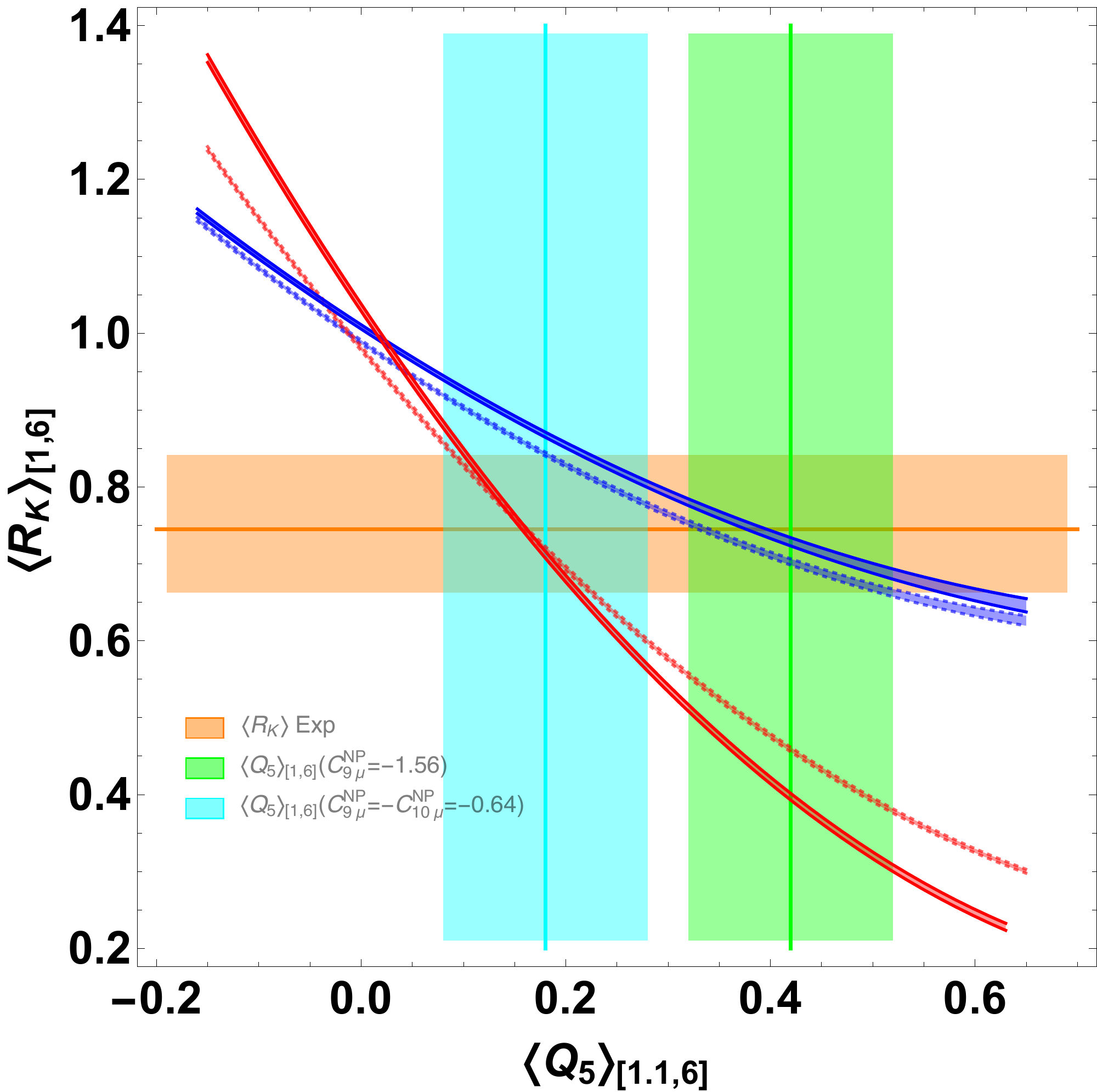}
\caption{$\langle R_K\rangle_{[1,6]}$ as a function of $\langle Q_5\rangle_{[1.1,6]}$ in the four of the scenarios analysed. The solid blue and solid red lines correspond to $C_{9,10}^{\rm U}=0$, while the dotted blue and dotted red lines have LFU contributions $\C{9}^{\rm U}=0.56$ and $\C{9}^{\rm U}=\C{10}^{\rm U}=-0.44$ respectively.}
\label{fig:plotRKQ5bin16LFU}
\end{figure}
Figure~\ref{fig:plotRKQ5bin16LFU} is a visual account of the decision tree discussed in the main text: in the case of an experimental determination of $\langle Q_5\rangle_{[1.1,6]}$ finding a value close to $0.4$ with enough precision (green band), only a solution involving $\C{9\mu}^\text{NP}$ (blue lines) can explain both $\langle Q_5\rangle_{[1.1,6]}$ and $\langle R_K\rangle_{[1,6]}$. However, this test has no discriminating power if $\langle Q_5\rangle_{[1.1,6]}$ is measured to be around 0.2 (blue band), since both $\C{9\mu}$ and $\C{9\mu}=-\C{10\mu}$ (red lines) scenarios could then explain $\langle Q_5\rangle_{[1.1,6]}$ and $\langle R_K\rangle_{[1,6]}$.
Another remarkable feature of this test is its robustness against its sensitivity to LFU-NP contributions. The solid curves in Figure~\ref{fig:plotRKQ5bin16LFU} correspond to $\langle R_K\rangle_{[1,6]}(\langle Q_5\rangle_{[1.1,6]})$ assuming there are no LFU contributions to the Wilson coefficients, while the dotted curves are realizations of the same functions but including contributions of the size suggested by our fits. As expected from the structure of the observables used in these tests, the inclusion of LFU-NP contributions barely induces corrections in the shapes of the curves.

\end{document}